\documentclass{article}
\usepackage{spconf,amsmath,graphicx}
\usepackage{array}
\usepackage{lipsum}
\usepackage{graphicx}
\usepackage{epstopdf}
\usepackage{url}

\title{Channel adversarial training for speaker verification and diarization}
\name{Chau Luu, Peter Bell, Steve Renals\thanks{Supported by an EPSRC iCASE studentship collaboration with the BBC.}}
\address{Centre for Speech Technology Research,  University of Edinburgh, UK\\
        \texttt{\{chau.luu, peter.bell, s.renals\}@ed.ac.uk} } 

\begin{document}
%
\maketitle
\begin{abstract}
  %

Previous work has encouraged domain-invariance in deep speaker embedding by adversarially classifying the dataset or labelled environment to which the generated features belong. We propose a training strategy which aims to produce features that are invariant at the granularity of the recording or channel, a finer grained objective than dataset- or environment-invariance. By training an adversary to predict whether pairs of same-speaker embeddings belong to the same recording in a Siamese fashion, learned features are discouraged from utilizing channel information that may be speaker discriminative during training. Experiments for verification on VoxCeleb and diarization and verification on CALLHOME show promising improvements over a strong baseline in addition to outperforming a dataset-adversarial model. The VoxCeleb model in particular performs well, achieving a $4\%$ relative improvement in EER over a Kaldi baseline, while using a similar architecture and less training data.
\end{abstract}


%
\section{Introduction}
\label{sec:intro}


Learning speaker discriminative features is an important approach to tasks such as
Speaker Verification (SV) and Speaker Diarization (SD). In recent years, using deep learning to extract speaker embeddings has become the state-of-the-art method for both tasks \cite{Snyder2018,Sell2018,Diez2019BayesianDiarization,Meng}, outperforming the well established i-vector technique \cite{Dehak2011}.


Although such embeddings have shown excellent performance for speaker verification, Probabilistic Linear Discriminant Analysis (PLDA) is often used to score the similarity between embeddings, rather than more direct measures such as cosine similarity or Euclidean distance. For i-vectors, the use of PLDA is motivated by the observation that often other unwanted sources of information, such as channel information, are present in these embeddings \cite{Dehak2011,Prince2007,Sell2014,GarciaRomero2011} -- thus training a separate model to disentangle these sources of information and extract only the speaker specific information has shown great benefit. The performance increase of PLDA when used with x-vectors \cite{Snyder2018,Sell2018}, suggests this source of unwanted variability is also present in deep embeddings.

This raises the question of whether disentangling channel information can be performed within the deep feature extractor, to either remove the need for PLDA, or to increase its effectiveness. The generation of channel-invariant features may be regarded as closely related to the production of domain-invariant features, for which adversarial training has emerged as a powerful approach to learning properties  such as domain-invariance in feature embeddings \cite{Ganin2015,shinohara2016adversarial,Shen2017}. 

Previous work in adversarial learning of speaker representation has encouraged domain invariance by having an adversary classify the dataset or labelled environment to which the generated features belong \cite{Meng,Tu2019}. However, this is a coarse modelling of the domains over which generated features are encouraged to be invariant. In the case of dataset adversarial training~\cite{Tu2019}, for instance, intra-dataset variation is not penalized, instead relying on the differences between datasets being enough to encourage meaningful invariance.

We aim to encourage invariance at the channel or recording level, without the need for labelled recordings, by training an adversary to predict whether pairs of same-speaker embeddings belong to the same recording. Since this recording-level adversarial penalty affects channel-related information, the approach encourages channel-invariant embeddings.

\begin{figure*}[tb]
  \includegraphics[width=\textwidth]{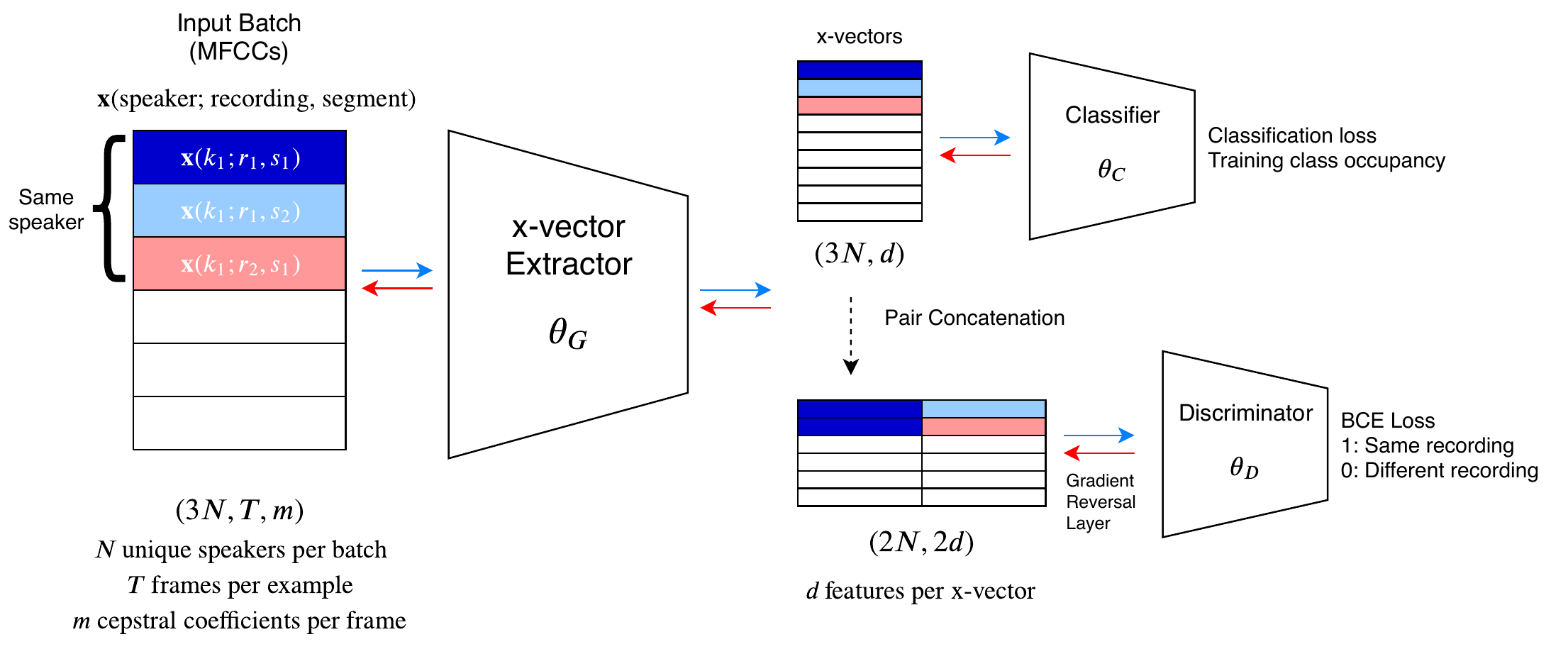}
  \caption{\label{fig:daxvector}The proposed architecture. The classifier is trained in the same way as the ordinary X-vector architecture, and the discriminator is trained on concatenated pairs of within-speaker pairs. The blue arrows represent the forward propagation, and the red arrows represent the backward propagation of gradients.}
\end{figure*}



Several researchers have performed related work using adversarial training to learn speaker embeddings.  Meng et al~\cite{Meng} used environment classification from a finite set of training environments and prediction of the signal to noise ratio of the input utterance as adversarial losses to a embedding generator for speaker verification. Tu et al~ \cite{Tu2019} aimed to make deep embeddings more amenable to PLDA by enforcing variational regularization to ensure a Gaussian distribution of representations. That work also incorporated an adversary to encourage domain invariance by having a discriminator perform a multi-class classification task on generated features, based on the dataset from which each sample originated. Bhattacharya et al~\cite{Bhattacharya2019} estimated deep speaker embeddings that were encouraged to be robust between single source and target domains using adversarial techniques.



Hsu et al~\cite{Hsu} proposed SiGAN, a Siamese architecture for upscaling faces using a generative network. By generating a pair of faces and ensuring that the performance of face verification was maintained, the generated faces were encouraged to be identity preserving. SiGAN is related to our work since it also focuses on  pairwise properties of generated features, which we use to encourage channel-invariance.

\section{Learning speaker embeddings}
\label{sec:speakerembeds}

We learn speaker embeddings by mapping a set of input frames of variable length $T$, $\textbf{X} = \{ \textbf{x}_1, ..., \textbf{x}_T\}$  to a fixed dimensional set of hidden features $\textbf{h}$ that represent the identity of a speaker, using a neural network parameterized by $\theta_G$. The x-vector neural network architecture \cite{Snyder2018} has been a particularly successful approach for this.

X-vectors are extracted from an intermediate layer of a network trained on classifying a set of speakers in a training set. From the input acoustic features of variable sequence length $\textbf{X}$, a series of Time Delay Neural Network (TDNN) layers (1-D convolutions in time) are applied sequentially, with each subsequent layer incorporating a larger temporal context. The output of the sequence of TDNN layers is pooled into a fixed dimension by taking the mean and variance of each unit of the frame-level output. This may then be projected into a smaller number of dimensions in order to extract the final speaker embedding. Up to this point in the network can be referred to as the embedding extractor, or \textit{generator}, parameterized by $\theta_G$.

Taking the speaker embeddings $\textbf{h}$ as input, a \textit{classifier} network learns to predict the input class. This network is trained with a loss function for multi-class classification, such cross entropy loss, $L_C$, and is parameterized by $\theta_C$.



\section{Channel Adversarial Training}
\label{sec:dat}

By training a discriminator, parameterized by $\theta_D$, to ascertain the domain of the generated features, an adversarial penalty is added to the overall loss function of a domain adversarial neural network (DANN)~\cite{Ganin2015,shinohara2016adversarial}:
\begin{equation}
    \mathcal{L}_{\text{DANN}}(\theta_g, \theta_c, \theta_d) = \mathcal{L}_C (\theta_c, \theta_g) - \lambda \mathcal{L}_D (\theta_d,\theta_g) \, ,
\end{equation}
where $\lambda$ is a controllable parameter to determine the weighting of this loss term. Allowing the adversary to act against the classifier is implemented via a gradient reversal layer between the generator and the discriminator.



In this work, the adversary classifies pairs of embeddings as being within-recording or not, thus penalizing the inclusion of channel information in the embeddings. This is implemented by attaching a discriminator with a gradient reversal layer that takes concatenated pairs of embeddings as input. This discriminator outputs a binary prediction.

This presents the question of which embeddings should be paired together to train the discriminator. Naively, one could select pairs randomly with a 50\% within-recording distribution. However, selecting pairs which do not have the same speaker may lead to the discriminator ascertaining the recording information based on the identities of the speakers, which is the opposite of the main training objective. As such, it is important that the discriminator only receives pairs of embeddings which belong to the same speaker.

This is achieved by selecting an anchor speaker $k$, and a random anchor utterance belonging to that speaker, $\textbf{x}(k; r_1, s_1)$: a segment $s_1$ belonging to a recording $r_1$. Within the same batch, another utterance from speaker $k$ is chosen from a separate segment $s_2$ of the same recording $r_1$, $\textbf{x}(k; r_1, s_2)$.  If such a segment does not exist, then $\textbf{x}(k; r_1, s_1)$ and $\textbf{x}(k; r_1, s_2)$ can be chosen by taking separated subsegments of a single utterance, with overlap minimized if possible. 

A second pair is constructed by choosing utterance $\textbf{x}(k; r_2, s_1)$, which can be a random utterance belonging to speaker $k$ which is not from recording $r_1$. If no out-of-recording utterances exist, or recording information is unknown, two subsegments of a single utterance can be taken to be the within-recording pair.

At the embedding stage, two pairs are concatenated, a within-recording pair $([\textbf{x}(k; r_1, s_1),\textbf{x}(k; r_1, s_2)])$ and an out-of-recording pair $([\textbf{x}(k; r_1, s_1),\textbf{x}(k; r_2, s_1)])$.

A batch for training is populated by selecting $N$ anchor speakers, and selecting the three segments  for each speaker, $(r_1, s_1)$,  $(r_1, s_2)$, and $(r_2, s_1)$. This results in an overall batch size of $3N$ for both the generator and the classifier, and an input batch size of $2N$ for the discriminator. The overall system is shown in Figure~\ref{fig:daxvector}, with colors to indicate the pattern of concatenation.

\section{Experimental setup}
\label{sec:exp}

We used the VoxCeleb\,1 \cite{Nagraniy2017} evaluation set and the CALLHOME corpus\footnote{\url{https://catalog.ldc.upenn.edu/LDC97S42}}  for our experiments. CALLHOME  is typically used for speaker diarization, so non-speaker-overlapping segments were extracted with minimum duration 0.5s according to the ground truth diarization,  in order to evaluate verification, selecting only pairs of segments occurring within the same recording.

The training data used was the same as in the Kaldi\footnote{\url{http://kaldi-asr.org}} recipes for VoxCeleb and CALLHOME. For training the VoxCeleb system, the VoxCeleb 2 \cite{Chung2018} corpus was augmented using background noises and room impulse responses as in the Kaldi recipe (although the Kaldi recipe also uses the training portion of VoxCeleb 1, which this work omits).

For  CALLHOME, the training data used was a combination of the NIST SRE 2004-2008 corpora, along with Switchboard 1, 2 and Cellular, all augmented in a similar fashion. Augmented versions were considered as different recordings.

\subsection{Baselines}
\label{subsec:baselines}

The network architecture for the generator closely follows Snyder et al~\cite{Snyder2018}, utilizing the same widths of temporal context at each layer, along with the choices for the number of hidden units  at each layer. Leaky ReLU and Batch Normalization were applied at each layer. 

Instead of using the stats pooling that the original architecture used, attentive stats pooling~\cite{Okabe2018} was used, with $128$ hidden units in the single attention head for the VoxCeleb system, and $64$ for the CALLHOME system. After pooling, the VoxCeleb system was projected to an embedding of size $512$, and  CALLHOME to a $128$-dimension embedding.

The classifier network was a single hidden layer feed forward network with $512$ hidden units for all models, projecting to the number of classes for each dataset. The classifier was trained using an additive margin softmax loss \cite{Wanga} using the recommended hyperparameters. All layers had a dropout schedule applied that started at $0$, rose to $0.2$ in the middle and dropped off to $0$ thereafter, similar to the Kaldi recipe.

Networks were trained on batches of utterances between $2-4$s in duration with batch size $400$, ensuring one example per speaker. Speakers were cycled in each batch to ensure a uniform distribution of speakers across training. The VoxCeleb system was trained for $100\,000$ batches and the CALLHOME for $25\,000$. SGD was used with learning rate $0.4$ and momentum $0.5$, with the learning rate halving at $60\%$ of the way through training, and halving for every $10\%$ thereafter.

For both VoxCeleb and CALLHOME there exist pretrained models in Kaldi, which were also used for benchmarking. Note that the Kaldi VoxCeleb model is trained using the VoxCeleb 1 training portion in addition to VoxCeleb 2.


\subsection{Acoustic features}
\label{subsec:feats}

For all experiments, $30$-dimensional MFCCs were extracted, with the standard $25$ms window and $10$ms step. Cepstral mean and variance normalization was applied to each utterance before training and only voiced frames were selected, judged by a simple energy based VAD system.

\subsection{Similarity scoring}

For both verification and diarization, either a cosine similarity or PLDA backend was used, utilizing length normalization for both. The PLDA model was trained on only the training data for that task, meaning either VoxCeleb 2 or the SRE-Switchboard combination. This differs particularly from some works on CALLHOME, which will train on some folds of the CALLHOME data, using the unseen folds for evaluation \cite{Lin2019,Zhang2018}. At no point have the models in this work been trained on any CALLHOME data.

\subsection{Diarization}

The diarization pipeline was as follows. From oracle speech activity marks, $1.5$s subsegments were extracted with a $0.75$s overlap. Speaker embeddings were extracted from each subsegment, normalized, and agglomerative hierarchical clustering was performed on the cosine similarity matrix. Cluster label overlaps were resolved by taking the mid-point of the overlap. Final diarization error rate was computed using \texttt{md-eval.pl}\footnote{\url{https://github.com/nryant/dscore/blob/master/scorelib/md-eval-22.pl}} with a forgiveness collar of $0.25$s.


\subsection{Adversarial Experiments}

To establish a baseline for other domain adversarial techniques, the CALLHOME model was also trained with a dataset-predicting adversary. The training data was split into three domain labels according to the dataset: SRE, Switchboard Cellular, or Switchboard. This adversarial discriminator was trained on the 3-class classification task on all embeddings in a batch using a cross entropy loss. This baseline was not possible with VoxCeleb due to the lack of domain label candidates.

The discriminator in all experiments was a simple feedforward network which had one hidden layer with $512$ units, outputting a single value for the within-recording prediction. For the channel-adversarial model, the size of the input was twice that of an embedding, so $1024$ for the VoxCeleb system and $256$ for the CALLHOME system. The gradient reversal layer $\lambda$ value was set to $1$.

\section{Results and Discussions}
\label{sec:results}

\begin{table}[tb]
  \centering
  \begin{tabular}{| m{3cm} || c c |}
  \hline
  & \multicolumn{2}{c|}{EER} \\
  \cline{2-3}
    & Cosine & PLDA \\ 
  \hline
   Baseline (Kaldi) & 9.77\% & 3.10\% \\
   Baseline (ours) & 5.94\% & 3.87\% \\
   Data-Tuned & 5.83\% & 3.92\%  \\
   Channel-Adversarial & \textbf{4.21\%} & \textbf{2.98\%}  \\
   \hline
  \end{tabular}
  \caption{\label{tab:vc_eer}EER values for the VoxCeleb 1 test set using cosine similarity or PLDA backend.}
\end{table}

\begin{table}[tb]
  \centering
  \begin{tabular}{| m{1.9cm} || c c c c|}
  \hline
  & \multicolumn{4}{c|}{EER} \\
  \cline{2-5}
  & \multicolumn{2}{c|}{All pairs} & \multicolumn{2}{c|}{Within-rec} \\
  \cline{2-5}
    & Cosine & PLDA & Cosine & PLDA \\ 
  \hline 
   BL (Kaldi) & 29.29\% & 19.06\% & 30.05\% & 23.16\% \\
   BL (ours) & \textbf{19.09\%} & 16.19\% & 28.51\% & 20.47\% \\
   Data-Tuned & 20.32\% & 17.75\% & 29.55\% & 22.43\% \\
   Dataset-Adv & 19.45\% & 16.30\% & 26.71\% & 20.55\% \\
   Channel-Adv & 21.11\% & \textbf{15.65\%} & \textbf{26.30\%} & \textbf{19.01\%} \\
   \hline
  \end{tabular}
  \caption{\label{tab:ch_eer}EER values for utterances from the CALLHOME dataset using cosine similarity or PLDA backend.}
\end{table}

\begin{table}[tb]
\centering
\begin{tabular}{| m{3cm} || c |}
\hline
& DER \\

\hline 
 Baseline (Kaldi) & 11.69\%  \\
 Baseline (ours) & 11.21\%  \\
 Dataset-Adv & 10.97\% \\
 Channel-Adv & \textbf{10.01\%}  \\
 \hline
\end{tabular}
\caption{\label{tab:ch_der}Diarization error rate on CALLHOME using a cosine similarity back-end.}
\end{table}

Table \ref{tab:vc_eer} shows speaker verification results on VoxCeleb for each model. When all components were trained from a random initialization, the channel-adversarial model did not converge. However, when the discriminator was added to an already converged baseline, the technique showed a marked improvement in performance, listed as `Channel-Adversarial' in the table. The `Data-Tuned' model is the control model that was trained from the same point as the channel-adversarial model but without an adversary -- this model never improves on the performance of the baseline. The improvement of our baseline over the Kaldi baseline for cosine similarity is likely due to  the use of attentive statistics pooling and the angular penalty softmax. The most comparable network architecture in the literature is that of Okabe et al~\cite{Okabe2018}, which achieves an EER of $3.8\%$ on VoxCeleb. In the recent VoxSRC\footnote{\url{http://www.robots.ox.ac.uk/~vgg/data/voxceleb/competition.html}} competition, much lower values for EER on VoxCeleb 1 were achieved ($< 2\%$), generally using much deeper models and also with higher dimension inputs. However, our results outperform others using small variations on the original x-vector architecture, in addition to outperforming some deeper models with more parameters \cite{Xie2019,Jung2019}.

Table \ref{tab:ch_eer} shows the verification performance of utterances from CALLHOME, for both within-recording pairs and across-recording pairs. Here, the channel-adversarial model with a PLDA backend produces the best EER in both scenarios. The adversarial models appear to perform better in general for within-recording pairs, with the channel-adversarial model performing the best once again, outperforming the dataset-adversarial model. Interestingly, the cosine similarity of the channel-adversarial model appears to degrade on the `all pairs' scenario.

Across all models, PLDA improves performance on verification, but the effectiveness of this improvement is somewhat unpredictable.

Table \ref{tab:ch_der} displays the diarization performance on CALLHOME using a cosine similarity backend, with the channel-adversarial model once again performing the best.


\section{Conclusions}
We have proposed a recording-level adversarial training strategy to reduce domain mismatch when estimating deep speaker embeddings. This is carried out by training an adversary to classify whether  pairs of embeddings belong to the same recording, thus penalising embeddings that contain channel information. Experimental results on VoxCeleb and CALLHOME show an improvement in performance by utilising this method over not only a standard baseline, but also an adversarial baseline which adversarially predicts training dataset occupancy.

\vfill\pagebreak

\bibliographystyle{IEEEbib}
\bibliography{strings,refs,references}

\end{document}